# СОСТОЯНИЕ МОБИЛИЗАЦИОННОЙ ГОТОВНОСТИ И ЧАСТОТНАЯ СТРУКТУРА ВАРИАБЕЛЬНОСТИ СЕРДЕЧНОГО РИТМА

В. Н. Мухин, В. М. Клименко

*ГУ НИИ экспериментальной медицины РАМН*

*e-mail: Valery.Mukhin@gmail.com*

*Взаимосвязь психических состояний с вариабельностью сердечного ритма показана во многих исследованиях. В настоящей работе изучена взаимосвязь психического состояния мобилизационной готовности с частотной структурой вариабельности сердечного ритма. Первый этап исследования проведён в трёх независимых группах (64, 39 и 19 человек). Методом факторного анализа периодограмм сердечного ритма обнаружено, что существует, по меньшей мере, два не известных ранее явления осцилляции сердечного ритма, кроме хорошо известных низкочастотных осцилляций, связанных с осцилляциями артериального давления, и дыхательной аритмии. Взаимосвязь амплитуды одной из этих осцилляций – имеющей период 3 кардиоинтервала – с уровнем мобилизационной готовности выявлена на втором этапе исследования в независимых группах (12 и 7 человек). Показана возможность оценки уровня мобилизационной готовности по амплитуде этой осцилляции.*

Эффективность деятельности во многом зависит от предшествующего ей и сопровождающего её особого психического состояния. Устоявшегося термина для обозначения этого состояния нет. В литературе его называют «готовность к соревнованию» [9], "оптимальное боевое состояние" [2],

"особое состояние спортсмена" [8] или «базовое активационное состояние» [5]. Мы будем использовать термин "мобилизационная готовность" [4].

Так как прогноз эффективности деятельности необходим, актуальна неспецифическая (не связанная с конкретным видом деятельности) объективная количественная диагностика мобилизационной готовности, но соответствующая методика отсутствует.

Однако, существуют предпосылки для её создания. Они заключаются в неспецифичности ряда особенностей состояния мобилизационной готовности: определённой внутренней картине (чувство полной гармонии душевных и физических сил) и максимальной функциональной интеграции на уровне центральной нервной системы.

С учётом этих предпосылок методика оценки мобилизационной готовности может быть разработана на базе анализа вариабельности сердечного ритма. Об этом свидетельствуют данные литературы: вариабельность сердечного ритма и, в особенности, мощность его высокочастотных осцилляций (0,15 – 0,4 Гц) зависит от психического состояния [см., напр., [3, 7, 11, 14, 16, 19]].

Несмотря на то, что об этой зависимости известно уже больше десяти лет, общепринятой методики диагностики психического состояния готовности до сих пор нет. Причина, по-видимому, в том, что параметры вариабельности сердечного ритма (связь которых с психическим состоянием обнаружена в указанных исследования) нерелевантны. Нерелевантность параметров связана с тем, что частотная структура вариабельности сердечного ритма и её физиологические механизмы изучены недостаточно [18].

Изучение структуры и механизмов затруднено тем, что в большинстве исследований при частотном анализе сердечного ритма имеется некоторое противоречие: серию временны́х интервалов (кардиоинтервалов) анализируют как функцию той же переменной (время). В связи с этим происходит избыточное дублирование информации о времени кардиоинтервалов по оси абсцисс. Преодолеть это противоречие можно, представляя сердечный ритм как функцию номера кардиоинтервала, то есть, как тахограмму. Несмотря на признание такого подхода в рекомендациях Европейского кардиологического общества и Североамериканского общества электрофизиологии [12], он почти не используется.

На базе такого подхода и с применением «разведочного анализа данных» может быть уточнена частотная структура вариабельности сердечного ритма (ВСР) и получены релевантные параметры её элементов. Это позволит изучить физиологические механизмы ВСР и разработать прикладные диагностические методики, в частности, методику количественной диагностики мобилизационной готовности.

Целью настоящего исследования явилось определение элементов частотной структуры вариабельности сердечного ритма, связанных с психическим состоянием мобилизационной готовности.

Для достижения цели требовалось решение двух задач: выявление элементов частотной структуры вариабельности сердечного ритма и изучение их связи с уровнем мобилизационной готовности.

**I. Выявление элементов частотной структуры вариабельности сердечного ритма.**

МЕТОДИКА

Регистрация сердечного ритма производилась путём записи и анализа ЭКГ в двенадцати общепринятых отведениях с помощью цифрового компьютерного электрокардиографа «Альтоника» с частотой дискретизации 1000 Гц. Автоматическая идентификация и классификация каждого комплекса QRS подтверждена визуально в соответствии с рекомендациями международной рабочей группы [12]. Участки записи, на которых выявлены нарушения ритма и проводимости, исключены из анализа. Составлены непрерывные равные между собой ряды из 300 RR-интервалов.

С помощью численного преобразования Фурье произведён частотный анализ этих рядов. Серии кардиоинтервалов проанализированы как функции номера интервала, то есть как тахограммы, поэтому абсцисса периодограмм выражена не в герцах, а в числах, обратных периоду, выраженному в кардиоинтервалах (ки$^{-1}$). В результате частотного анализа определены 150 гармоник (N/2) в диапазоне от 0 до 0,5 ки$^{-1}$ с шагом в 1/300 ки$^{-1}$. Значения периодограмм на каждой частоте логарифмически преобразованы, что позволило нормализовать их распределение.

С целью выявления элементов частотной структуры сердечного ритма произведён факторный анализ периодограмм (разведочный анализ данных). Использован метод главных компонент с вращением «варимакс». Далее – произведён графический анализ полученных компонент. Выявление элементов частотной структуры периодограмм сердечного ритма производилось в трёх независимых группах: 64, 39 и 19 человек; во второй и третьей группах – для проверки результатов первой группы, что обеспечило бо́льшую надёжность результатов. Взаимосвязь между факторными компонентами в разных группах подтверждена статистически (коэффициентами корреляции Пирсона).

РЕЗУЛЬТАТЫ

Результаты факторного анализа оказались аналогичными в каждой из трёх групп: диаграммы факторных нагрузок четырёх первых факторов имели форму волны (рис. 1). Волны соответствующих друг другу факторов в разных группах располагались на одних и тех же частотах: первые факторы 0,27 – 0,50 ки$^{-1}$, вторые – 0,09 – 0,28 ки$^{-1}$, третьи – 0,22 – 0,27 ки$^{-1}$ и четвёртые – 0,09 – 0,17 ки$^{-1}$. Их вершины также лежали приблизительно на одних и тех же частотах: первые факторы – 0,33 ки$^{-1}$, вторые – 0,19 ки$^{-1}$,

третьи – 0,26 ки⁻¹ и четвёртые – 0,13 ки⁻¹. В связи с этим мы предположили, что волны факторных нагрузок (ВФН) в разных группах соответствуют друг другу. Результаты корреляционного анализа нагрузок соответствующих факторов в разных группах (см. таблицу) подтвердили это.

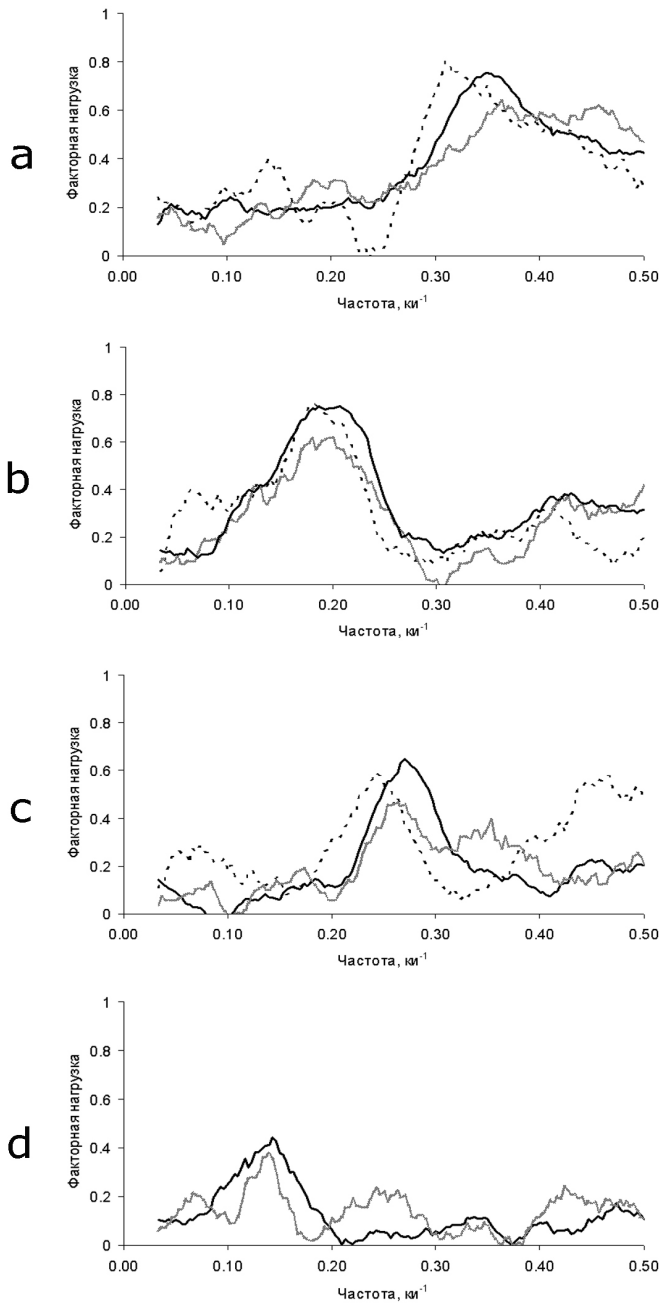

Рис. 1. Сравнительные диаграммы факторных нагрузок. Применено сглаживание скользящим средним арифметическим по 9 точкам. a – первые факторы, b – вторые факторы, c – третьи факторы, d – четвёртые факторы. Сплошная линия – первая группа, пунктирная линия – вторая группа, серая линия – третья группа. Четвёртый фактор во второй группе не обозначен, так как он не имел форму волны.

**Корреляции между соответствующими друг другу факторами, полученными в результате факторного анализа периодограмм сердечного ритма в разных группах**

| Пары групп | Фактор 1 | Фактор 2 | Фактор 3 | Фактор 4 |
|---|---|---|---|---|
| Группа 1 и группа 2 | 0,81 | 0,79 | 0,33 | |
| Группа 1 и группа 3 | 0,86 | 0,90 | 0,73 | 0,43 |
| Группа 2 и группа 3 | 0,62 | 0,71 | 0,26 | |

*Примечание. Показаны только статистически значимые коэффициенты корреляции (p≤0,05).*

То, что на периодограммах сердечного ритма можно наблюдать пики в частотных пределах, соответствующих ВФН (рис. 2), позволило убедиться в обусловленности каждой ВФН осцилляцией сердечного ритма на определённой частоте.

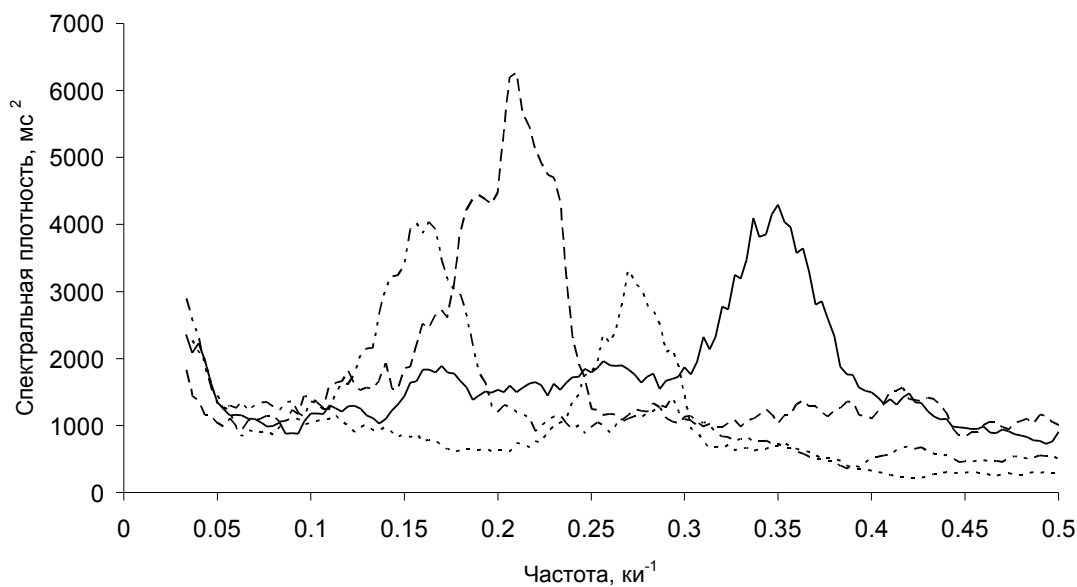

*Рис. 2. Примеры спектрограмм сердечного ритма, на которых хорошо выражены пики на тех же частотах, что и волны на диаграммах факторных нагрузок. Применено сглаживание скользящим средним арифметическим по 9 точкам. Сплошная линия – спектрограмма имеет пик на частотах ВФН первого фактора, короткий пунктир –*

*спектрограмма имеет пик на частотах ВФН третьего фактора, длинный пунктир – спектрограмма имеет пик на частотах ВФН второго фактора, штрихпунктирная линия – спектрограмма имеет пик на частотах ВФН четвёртого фактора.*

Иногда существование осцилляций на этих частотах обнаруживается и непосредственно на кардиоинтервалограммах (рис. 3).

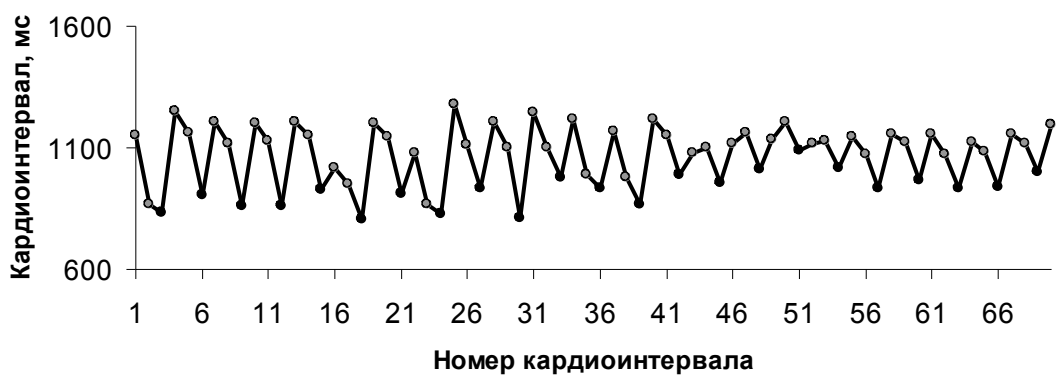

***Рис. 3.*** *Пример тахограммы сердечного ритма с хорошо выраженной модуляцией, имеющей период 3 кардиоинтервала. Чтобы подчеркнуть периодичность, каждый третий кардиоинтервал обозначен точкой чёрного цвета.*

Показанная на рисунке 3 осцилляция с частотой 0,33 ки$^{-1}$ (периодом 3 кардиоинтервала) обладает наибольшим диапазоном изменения своей амплитуды, так как соответствующий ей фактор имеет наибольшее собственное значение в каждой из трёх групп (до 24%).

Аналогичный факторный анализ периодограмм сердечного ритма всех групп вместе позволил получить более надёжные частотные параметры ВФН (рис. 4), так как выборка в этом случае была больше (N=122).

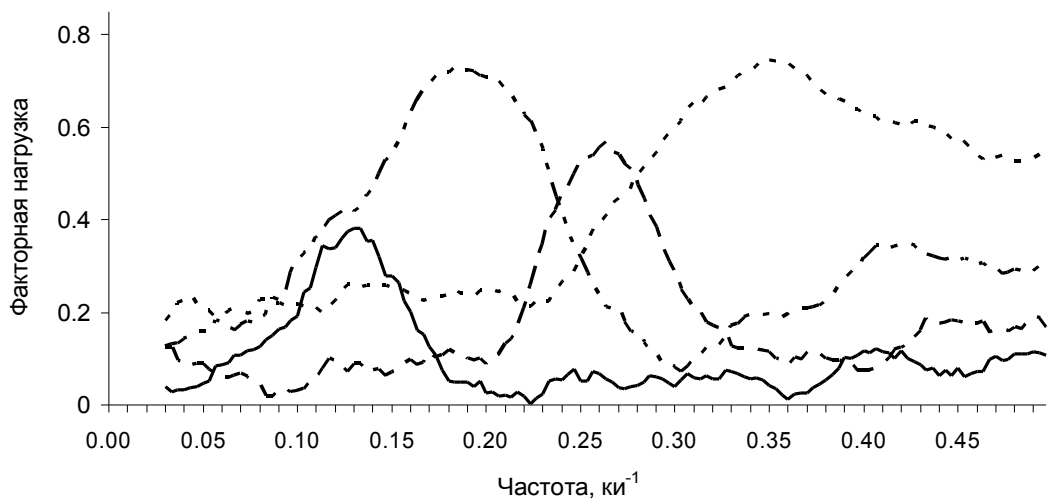

***Рис. 4.*** *Диаграммы четырёх первых факторов. Все группы вместе (N=122). Применено сглаживание скользящим средним арифметическим по 9 точкам. Короткий пунктир – первый фактор, штрихпунктирная линия – второй фактор, длинный пунктир – третий фактор, сплошная линия – четвёртый фактор.*

ОБСУЖДЕНИЕ

Как свидетельствуют результаты исследования, в частотную структуру ВСР входят, по меньшей мере, 4 осцилляции сердечного ритма.

Две из них (соответствующие четвёртому и второму факторам) хорошо известны; так, осцилляция, обусловливающая ВНФ четвёртого фактора, соответствует известным из литературы медленным волнам первого порядка (низкочастотные волны, LF), связанным с волнами артериального давления (волны Майера) и с терморегуляцией [13]. По результатам факторного анализа она имеет наиболее вероятный частотный диапазон – 0,09 - 0,17 ки$^{-1}$. Если учесть, что в состоянии относительного покоя (в котором производилась запись сердечного ритма) частота сердечных сокращений в норме 1,0 – 1,5 Гц, то получается, что нижний частотный предел этой осцилляции во временном представлении не может быть менее 0,090 ки$^{-1}$ · 1 Гц = 0,090 Гц, а верхний частотный предел не может быть более 0,17 ки$^{-1}$ · 1,5 Гц = 0,255 Гц. Из литературы следует, что в покое на этих частотах (0,090 – 0,255 Гц) с наибольшей вероятностью наблюдаются низкочастотные осцилляции (их пик – 0,1 – 0,12 Гц) [1, 10, 17].

Осцилляция, обусловливающая ВФН второго фактора, соответствует дыхательным волнам, что подтверждается совпадением частоты этой осцилляции (0,19 ки$^{-1}$) с частотой дыхания в покое по отношению к сердечному ритму. Другое подтверждение – в том, что частотные границы ВФН этого фактора, экстраполированные (как в предыдущем случае) во временные единицы (0,09 – 0,42 Гц), соответствуют нормальной частоте дыхания в покое (0,18 – 0,33 Гц).

Сведения об осцилляциях, соответствующих ВФН первого и третьего факторов (с наиболее вероятными частотами 0,33 ки$^{-1}$ и 0,26 ки$^{-1}$) нами в литературе не найдены; их механизмы нам неизвестны.

**II. Связь элементов частотной структуры сердечного ритма с уровнем мобилизационной готовности**.

МЕТОДИКА

Были созданы две группы (из числа участвовавших в предыдущей части исследования): первая – 12 человек – для обнаружения связи, а вторая – 7 человек – для подтверждения результата.

В каждой из этих групп в дополнение к регистрации сердечного ритма был определён уровень мобилизационной готовности. Использован русскоязычный вариант психологического опросника POMS [15], разработанный в Санкт-Петербургском НИИ физической культуры под руководством профессора П. В. Бундзена [6]. Одна из шкал опросника, «психическая сила» (V), характеризует уровень мобилизационной готовности: испытуемому предлагается оценить, насколько он оживлён, активен, энергичен, бодр, воодушевлён, полон сил. Надёжность этой шкалы проверена путём определения альфы Кронбаха на выборке из 59 испытуемых.

Далее, отдельно в каждой группе, изучена связь элементов частотной структуры ВСР, выявленных в предыдущей части исследования, с уровнем мобилизационной готовности. Поскольку определённой гипотезы о такой связи не было, также применены методики «разведочного анализа данных». Первая – анализ корреляционной матрицы между шкалой «психическая сила» и частотами периодограммы сердечного ритма. Она позволила обнаружить те частоты, на которых амплитуда осцилляций сердечного ритма связана с уровнем мобилизационной готовности. Вторая методика – построение множественной регрессионной модели шкалы V опросника путём автоматизированного пошагового подбора переменных. Она позволила подтвердить обнаруженную взаимосвязь и выявить возможность диагностики уровня мобилизационной готовности путём анализа периодограмм сердечного ритма.

## РЕЗУЛЬТАТЫ

Выявлена положительная взаимосвязь между значениями шкалы V психологического опросника и значениями периодограммы сердечного ритма на частоте 0,333 ки$^{-1}$ в обеих группах: в первой (R=0,68; p=0,016) и во второй (R=0,79; p=0,034). Надёжность этой шкалы не вызывает сомнений, так как альфа Кронбаха для неё в нашем исследовании составила 0,859. Значит, мы можем полагать, что уровень мобилизационной готовности положительно связан с амплитудой осцилляции сердечного ритма на частоте 0,333 ки$^{-1}$ (период 3 кардиоинтервала), выявленной в предыдущей части исследования.

Возможность оценки мобилизационной готовности по амплитуде этой осцилляции показал множественный регрессионный анализ. Получены регрессионные модели шкалы V в обеих группах на базе одних и тех же частот периодограммы: 0,333 ки$^{-1}$, 0,327 ки$^{-1}$, 0,313 ки$^{-1}$. Величина коэффициента R-квадрат позволяет считать диагностическую способность моделей достаточно высокой: в первой группе – 0,789 (скорректированный – 0,742; p<0,00001), во второй группе – 0,816 (скорректированный – 0,724; p<0,00001). B-коэффициенты моделей в обеих группах статистически достоверны (p<0,05). Распределение остатков в обеих группах – нормальное. Наконец, визуальный анализ скаттергамм «наблюдения»-«предсказания» (рис. 5) подтвердил хорошую предсказательную способность моделей: точки располагаются равномерно, в виде вытянутого под углом облака.

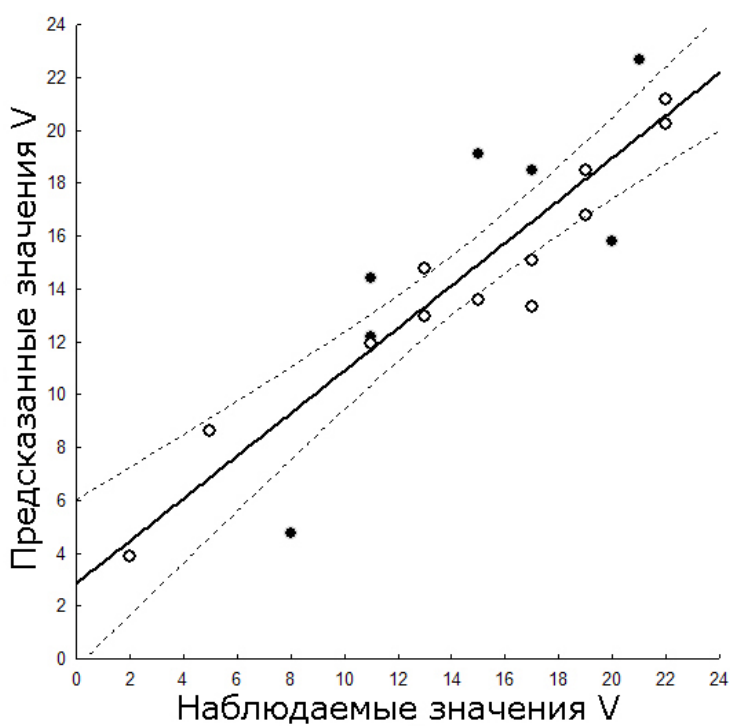

***Рис. 5****. Диаграмма "наблюдения"-"предсказания" модели шкалы V (психическая сила). Кружочки – первая группа, чёрные точки – вторая группа.*

Принимая во внимание, что модели, по существу, представляют собой разность между значениями периодограммы на частоте 0,333 ки$^{-1}$ (B-коэффициент положителен) и на частоте 0,313 ки$^{-1}$ (B-коэффициент отрицателен), можно заключить, что они дают представление об амплитуде осцилляции сердечного ритма с периодом 3 кардиоинтервала. При этом в модель включены те частоты периодограммы, на которых влияние других осцилляций сердечного ритма минимально (минимальны нагрузки ВФН других факторов). Значит, модель имеет смысл в рамках выявленной в предыдущей части исследования частотной структуры вариабельности сердечного ритма.

## ОБСУЖДЕНИЕ

Данные литературы свидетельствуют, что психические состояния связаны главным образом с модуляцией сердечного ритма в частотном диапазоне 0,15 – 0,40 Гц (высокие частоты, HF) [11, 14, 16, 19]. Результаты исследования показали, что психическое состояние мобилизационной готовности связано лишь с одной из нескольких осцилляций сердечного ритма, наблюдающихся в этом частотном диапазоне. Значит, мощность периодограммы сердечного ритма в этом диапазоне не может служить надёжным диагностическим показателем мобилизационной готовности. Простое изменение границ зоны также не ведёт к повышению надёжности, так как частотные зоны «соседних» осцилляций частично перекрываются (что следует из результатов исследования) и их влияние на общую мощность спектра в зоне с любыми границами неизбежно.

Но математическая модель позволяет, как показали результаты исследования, «вычесть» влияние «соседних» осцилляций, и, как следствие, получить надёжную оценку уровня мобилизационной готовности по амплитуде осцилляции сердечного ритма с периодом 3 кардиоинтервала.

## ЗАКЛЮЧЕНИЕ

Частотная структура вариабельности сердечного ритма более сложна, чем принято считать в настоящее время. Обнаружено две, не описанных ранее в литературе, осцилляции сердечного ритма, имеющих период 3 и 4 кардиоинтервала. Амплитуда одной из них – период 3 кардиоинтервала – положительно связана с уровнем мобилизационной готовности. Оценка мобилизационной готовности по амплитуде этой осцилляции возможна.

## СПИСОК ЛИТЕРАТУРЫ

[1] Аксёнов В. В. Методические основы кибернетического анализа сердечного ритма. В сб.: Ритм сердца у спортсменов. Под ред. Р. М. Баевского, Р. Е. Мотылянской. Физкультура и спорт, М. 1986. С. 36.

[2] Алексеев А. В. Психофункциональный тест – способ оценки психической подготовленности спортсменов. М. 1979.

[3] Бундзен П. В., Мухин В. Н. Использование анализа вариабельности сердечного ритма в оценке психофизического потенциала спортсменов-учащихся училищ олимпийского резерва. В сб.: Итоги II спартакиады "Спортивный потенциал России". Орёл. 2004. С. 195-217.

[4] Генов Ф. Психологические особенности мобилизационной готовности спортсменов. Физкультура и спорт, М. 1971.

[5] Ильин В. П. Психофизиологические состояния человека. М. 2005. 410 с.

[6] Исаков В. А. Физиологические механизмы стресслимитирующего эффекта ментальной релаксации. Дис. ...канд. мед. наук. Самара. 2002.

[7] Машин В. А., Машина М. Н. Анализ вариабельности ритма сердца при негативных функциональных состояниях в ходе сеансов психологической релаксации. Физиология человека. 26 (4) : 48–54. 2000.

[8] Некрасов В. П., Худадов Н. А., Пиккенхайн Л., Фрестер Р. Психорегуляция в подготовке спортсменов. Физкультура и спорт, М. 1985.

[9] Пуни А. Ц. Волевая подготовка в спорте. М. 1969.

[10] Akselrod S., Gordon D., Ubel F. A., Shannon D. C., Barger A. C., Cohen R.J. Power spectrum analysis of heart rate fluctuation: a quantitative probe of beat to beat cardiovascular control. Science. 213 : 220–222. 1981.

[11] Friedman B. H., Thayer J. F. Autonomic balance revisited: panic anxiety and heart rate variability. J Psychosom Res. 44 (1) : 133–151. 1998.

[12] Heart Rate Variability. Standards of Measurement, Physiological Interpretation, and Clinical Use. Circulation. 93 : 1043–1065. 1996.

[13] Hyndman B. W., Kitney R. I., Sayers B. McA. Spontaneous Rhythms in Physiological Control Systems. Nature. 233 : 339–341. 1971.

[14] Laskar M. S., Iwamoto M., Toibana N., Morie T., Wakui T., Harada N. Heart rate variability in response to psychological test in hand-arm vibration syndrome patients assessed by frequency domain analysis. Ind Health. 37 : 382–389. 1999.

[15] McNair D. M., Lorr M., Droppleman L. F. Manual: Profile of Mood States. CA: "Educational and Industrial Testing Service", San Diego. 1971.

[16] Mezzacappa E., Tremblay R. E., Kindlon D., Saul J. P., Arseneault L., Seguin J., Pihl R. O., Earls F. Anxiety, antisocial behavior, and heart rate regulation in adolescent males. J Child Psychol Psychiatr. 38 : 457-469. 1997.

[17] Sayers B. M. Analysis of heart rate variability. Ergonomics. 16 : 17–32. 1973.

[18] Taylor J. A., Studinger P. Point:Counterpoint: Cardiovascular variability is/is not an index of autonomic control of circulation. J Appl Physiol. 101 : 678–681. 2006.

[19] Valkonen-Korhonen M., Tarvainen M. P., Ranta-Aho P., Karjalainen P. A., Partanen J., Karhu J., Lehtonen J. Heart rate variability in acute psychosis. Psychophysiology. 40 : 716-726. 2003.